\documentclass[epsfig,twocolumn]{revtex4}
\usepackage{graphicx}
\usepackage{amsmath}
\usepackage{latexsym}
\usepackage{amsfonts}
\usepackage{amssymb}
\usepackage{array}
\usepackage{color}
\usepackage{subfigure}
\usepackage{verbatim}
\newcommand{\ket}[1]{\left\vert#1\right\rangle}
\newcommand{\bra}[1]{\left\langle#1\right\vert}

\newcommand{\braket}[2]{\left\langle#1|#2\right\rangle}

\begin{document}
\newcommand{\Q}[1]{{\color{red}#1}}
\newcommand{\B}[1]{{\color{blue}#1}}
\newcommand{\blue}[1]{{\color{blue}#1}}
\newcommand{\red}[1]{{\color{red}#1}}
\newcommand{\green}[1]{{\color{green}#1}}
\newcommand{\Change}[1]{{\color{green}#1}}

\title{Hybrid Rabi interaction with traveling states of light}

\author{Kimin Park}
\affiliation{Department of Optics, Palack\'y University, 17. listopadu 1192/12, 771 46 Olomouc, Czech Republic}
\affiliation{Center for Macroscopic Quantum States (bigQ), Department of Physics,
Technical University of Denmark, Building 307, Fysikvej, 2800 Kgs. Lyngby, Denmark}

\author{Julien Laurat}
\affiliation{Laboratoire Kastler Brossel, Sorbonne Université, CNRS, ENS-Université PSL, Collège de France, 4 place Jussieu, 75005 Paris, France}

\author{Radim Filip}
\affiliation{Department of Optics, Palack\'y University, 17. listopadu 1192/12, 771 46 Olomouc, Czech Republic }
\date{\today}

\begin{abstract}
Hybrid interactions between light and two-level systems and their nonlinear nature are crucial components of advanced quantum information processing and quantum networks. 
Rabi interaction  exhibits the hybrid nonlinear nature, but its implementation is challenging at optical frequencies where the rotating wave approximation (RWA) is valid.  
Here, we propose a setup to conditionally induce Rabi interaction  between discrete variable and continuous variable of traveling beams of light.  
We show that our  scheme can generate  Rabi interaction on weak states of light,  where signatures of the nonlinear quantum effects are preserved for typical experimental losses.   These results prove that a hybrid Rabi interaction can be realized in all-optical setups, and open a way to experimental investigations of nonlinear quantum optics beyond RWA. 
\end{abstract}

\maketitle
 
\section{Introduction}

Quantum technologies and their advanced applications heavily depend on the ability to implement strong nonlinear operations at different experimental platforms~\cite{ChangNatPho2014,Kurizki2015Hybrid}. Any interaction between qubits can be universally decomposed to a circuit of basic logic gates, such as controlled-NOT gates~\cite{NielsenBook}. However, for a harmonic oscillator in an infinite-dimensional Hilbert space, such a universal decomposition of arbitrary interactions is non-trivial. For optical modes,  decomposition can be typically done by first expressing an interaction as a sequence of ladder operators $\hat{a}$ and $\hat{a}^{\dagger}$ \cite{ZavattaScience2004,OurjoumtsevScience2006,Neergaard-NielsenPRL2006,ParigiScience2007,TakahashiNatPho2010, TrepsQIM2017Subtraction, RaTreps2019NGMulti}.  Recently, non-Gaussian quantum states and operations   conditionally achieved by photon subtractions and additions have been proposed to achieve noiseless amplifier~\cite{ZavattaNatPho2011Amplifier,UsugaNatPhy2010NoiseProbConcentPhase}, entangle macroscopic states~\cite{Biago2018MacroEnt} to apply them in teleportation \cite{TakedaNature2013Teleportation, LvovskyPRL2017HybTelep, TakedaPRL2015Swap}, remote state preparation \cite{MorinNat2014} and quantum steering \cite{LeJeannicOptica2018, CavaillesPRL2018HybridSteering}. A conditional superposition of ladder operators can experimentally emulate  nonlinearity for weak states of light \cite{CostanzoPRL2017}. Alternatively, a conditionally applied sequence of monomials of position quadrature operator $\hat{X}$ can conditionally implement a single-mode unitary operator of the form $\exp[i t V(\hat{X})]$ with a strength $t$, where $V(\hat{X})$ is any nonlinear  single-variable function of  $\hat{X}=(\hat{a}+\hat{a}^\dagger)/\sqrt{2}$ \cite{ParkPRA2014, ArzaniPRA2017}.  This approach can be extended to simulate the superpositions of two unitary operators \cite{ParkPRA2015SuperpositionUnitary} and challenging quantum model of  optomechanical interactions in the membrane-in-the-middle setup~\cite{ParkPRA2015Optomechanics}. For deterministic approaches, using repeated applications of highly nonlinear cubic phase gates, any nonlinear interaction can be achieved  in principle  \cite{SefiPRL2011,MarekPRA2018Phase}. However, the cubic phase gate is classically unstable with unbounded energy eigenstates, and as such, experiments are challenging~\cite{MarshallPRA2015Repeat, MiyataPRA2016cubicadaptive}. Many other types of nonlinearities are also achieved deterministically using various physical systems~\cite{ChangNatPho2014}, such as optical cavities~\cite{optcav2005,optcav2008,optcav2008science, optcav2011science, optcav2012}, atomic ensembles~\cite{AtomEnsemblesRMP2010}, strong Rydberg atoms interaction~\cite{RydAtomPRL2008}, or many-body nonlinear media~\cite{polaritonNatPhy2006,plasmonsNatPhy2007,NanowireNat2007, PhotonicCrystalNatCom2014,ChiralNat2017}.

Rabi interaction (RI), the direct coupling  in the Rabi model between the quadrature variable of a quantized field and an atomic polarization in ultrastrong coupling regime~\cite{RabiHist,KockumNatRevPhy2019, Forn-DiazRMP2019Rabi},  is naturally present at low frequencies in mechanical and microwave systems~\cite{ChangRMP2018AtomPhotonLattices,RabiOperationIon,RabiOperationIon2,RabiOperationIon3,Lv2017RabiSimulation,RabiOperationCQED,RabiOperationCQED2,RabiOperationCQED3,RabiOperationCQED4,RabiOperationCQED5,LangfordNatCom2017CQED,TodorovPRL2010USCpolaritondot}.   Recently, a sequence of RIs between a qubit and an oscillator has been proposed to induce the deterministic Kerr, cubic or arbitrary-order nonlinear phase gates \cite{ParkKerr2017, ParkCubic2017}. It was also studied for exhibition of universal phase transition properties \cite{HwangPRL2015,Peng2019SPT,Felicetti2019SpectralUSC}, multi-photon exchange~\cite{GarzianoPRA2015}, stimulated emission \cite{NoriNJP2017RabiEmission}, microwave-to-optical conversion \cite{Lambert2019conversion}, generation of non-Gaussian states ~\cite{FluhmannHome2019NatureGKP, Devorat2019GKP, WangScience2016Cat}, and decomposition of arbitrary unitary dynamics~\cite{ParkPRA2016Rabi}. This power enabling synthesis of various types of nonlinearity and observation of consequent nonlinear properties  is the reason why achieving RI on continuous variable platforms is important.  However,  the RI is reduced to the Jaynes-Cummings (JC) interaction under the rotating-wave approximation (RWA) which averages out off-resonant terms and keeps only the energy-conserving ones  \cite{Louisell1973RWA}. This approximation fits very well for optical frequencies experiments, and therefore a true RI is challenging to reach for light~\cite{Gambino2014Rabi,Cacciola2014Rabi}. Until now, the only accessible methods for optical implementation are digital simulation~\cite{RabiOperationIon,RabiOperationIon2,RabiOperationIon3,Lv2017RabiSimulation,LangfordNatCom2017CQED} based on stroboscopic application of frequency-detuned JC interaction, and analog simulation \cite{BraumullerNatCom2017AnalogRabi, MarkovicPRL2018, Peterson2019USCcavityResonator} based on orthogonal driving. This method is suitable for trapped ion experiments, but is still challenging at high optical frequencies.

Hybrid quantum optics aim at combining the advantages of discrete-variable (DV) and continuous-variable (CV) quantum optics, and thus reaching a regime beyond both platforms to overcome their individual limitations and reach full control of quantum systems~\cite{ULA2015Hybrid,Takeda2016hybrid}. Examples are quantum teleportation of a DV using CV protocol ~\cite{TakedaNature2013Teleportation} or the teleportation of CV qubit to qubit~\cite{LvovskyPRL2017HybTelep}, quantum repeater using hybrid protocol~\cite{AndersenPRL2010HybSwap} or building on-chip integrated circuits~\cite{ElshaariNatComm2017HybridCircuit}. Hybrid optical states have been generated experimentally to entangle the DV and CV~\cite{JeongNat2014,AndersenPRA2013,MorinNat2014,PvL2011hybrid, TakedaPRL2015Swap,LeJeannicOptica2018,CavaillesPRL2018HybridSteering,MinzioniJO2019,HuangNJP2019HybridEntanglement,CavaillesQIM2017Steering, VogelPRL2017Nonclassicality}.
Therefore, a natural question arises about whether and up to what extent we can induce the nonlinear effects of RI all-optically.

In this work, we attempt to tackle this open question and propose a feasible scheme to engineer a quantum RI for weak states of light and simulate the nonlinear effects beyond the RWA using hybrid quantum optical toolbox.  In Sec. \ref{secII}, we introduce the methods to achieve the approximations of RI in the lowest order expansions. In Sec. \ref{secIII}, we quantify the entangling process  of the scheme between CV and DV degrees of freedom to prove its power to reach beyond the RWA. In Sec. \ref{secIV}, a remote preparation of a superposition of displacement operations is provided as a direct evidence of the dynamics beyond the JC type.   For this purpose, we analyze the important aspects of the RI for experimental verification using homodyne quantum tomography. In Sec. \ref{secV}, we conclude. This work is intended as a proposal for a first proof-of-principle experiment.  

\section{Optical implementation of finite-order expansions}
\label{secII}

Quantum Rabi model describes an evolution of a dipole-field system under a Hamiltonian $\hat{H}=\hat{H}_{0}+\hat{H}_\mathrm{int}$, a sum of the self-energy $\hat{H}_{0}=\hat{\sigma}_z/2+\hat{n}$ and the interaction Hamiltonian $\hat{H}_\mathrm{int}=\hat{\sigma}_h \hat{X}_\theta$. 
Here, the Pauli matrices $\hat{\sigma}_{h=x,y,z}$ represent atomic polarization of a two-level system, and the photon number $\hat{n}=\hat{a}^\dagger \hat{a}$  and  a quadrature operator $\hat{X}_\theta=(\hat{a}e^{-i\theta}+\hat{a}^\dagger e^{i\theta} )/\sqrt{2}$ at phase $\theta$ is of an harmonic oscillator.  
The position quadrature operator, i.e. $\hat{X}=\hat{X}_{\theta=0}$ resides in an infinite dimensional space represented by a position quadrature eigenbasis $\{\ket{x}\}$ satisfying $\hat{X}\ket{x}=x \ket{x}$ with a continuous  spectrum of eigenvalues $x$. 
The Pauli matrix $\hat{\sigma}_x$ in a two-dimensional  space with Pauli index $h=x$ without loss of generality, on the other hand, possesses a discrete eigenstate spectrum, i.e. atomic basis  $\{\ket{e},\ket{g}\}$.
This model possesses an inherent hybrid nature imposed by the asymmetric dimensions  of the two involved operators $\hat{X}$ and $\hat{\sigma}_x$. 
% Simultaneously, it is nonlinear interaction due to saturation of two-level system. 

 The simulation of this model with two optical oscillators therefore requires confinement of the infinite Fock space of an oscillator into the two-dimensional qubit space with discrete eigenstates.   We use now a freedom of the association of the optical basis with the atomic basis.   For the simulation in hybrid quantum optical systems, the optical qubit may flexibly represent a virtual atomic basis  $\{\ket{e},\ket{g}\}$ or any other decomposition of the qubit space. 
 We choose the simplest association  $\ket{e}=\ket{1}_d$ and $\ket{g}=\ket{0}_d$ throughout this work where $\ket{n}$ is a Fock state with a photon number $n$ of optical mode used to represent two-level system. 
 Then the qubit excitation is changed into optical excitation  and the self-energy becomes $\hat{H}_{0}=\hat{n}_1+\hat{n}_2$, which has only trivial effects such as rotation of the phase space axes. Therefore this term can be omitted in the interaction picture, where the dynamics is solely described by a unitary evolution operation $\hat{U}_\mathrm{Rabi}(t)=\exp[i t\hat{H}_\mathrm{int}]$ with a strength-time product $t$,  the only relevant term for quantum processing with continuous variables. 
 
 \subsection{First-order approximation}
  The evolution under a RI incurs highly nonlinear effects on the involved qubit and oscillator due to the saturation of two-level system and a combination of inherent quantum nonlinearity. 
  Moreover, the interaction $\hat{H}_\mathrm{int}$ is beyond RWA and at high frequencies it converges to the JC interaction. 
As an example of its nonlinear effects, an infinite-order polynomial operation $\hat{O}_R=\cos[t \hat{X}_\theta]+i\delta\sin[t\hat{X}_\theta]$ with an arbitrary complex number $\delta$ is applied to the oscillator after a projective detection on the qubit mode, capable to induce a strong nonlinearity conditionally~\cite{ParkPRA2016Rabi}.   Therefore, a question should be addressed whether RI can be achieved by a purely optical setup devised to implement a hybrid entangling gate between a discrete variable (DV) and a continuous variable (CV) degrees of freedom.  The RI conserves the expectation values of the local variables $\hat{X}$ and $\hat{\sigma}_{x}$ while their respective conjugate variables are shifted,  capable of preparing the superposition of displacement operation $\exp[\pm i t \hat{X}]$ in conjugate momentum $\hat{P}=\hat{X}_{\pi/2}$. The optical simulation can therefore be accomplished by a coherent control of the direction of a displacement  (the sign of $t$) by discrete optical control qubit states  $\{\ket{0},\ket{1}\}$.   
A general evolution  by a RI of CV mode in an arbitrary state  $\ket{\psi}$ and DV mode in an arbitrary qubit state  $\ket{\phi_0}=c_+\ket{+}+c_-\ket{-}$ with $c_+,c_-\in \mathbb{C}$  can be decomposed into the  eigenstates of the Pauli operator $\hat{\sigma}_x\ket{\pm}=\pm \ket{\pm}$ as
\begin{align}
\exp[it \hat{\sigma}_x \hat{X}]\ket{\phi_0}\ket{\psi}=c_+\ket{+}e^{it \hat{X}}\ket{\psi}+c_-\ket{-}e^{-it \hat{X}}\ket{\psi}.
\end{align}
The simplest implementation is the first order expansion of a weak RI $\hat{U}_\mathrm{Rabi}(t)\approx 1+i t\hat{\sigma}_x \hat{X}=\hat{U}^{(1)}(t)$ where $\hat{U}^{(j)}(t)= \sum_{k=0}^j (it \hat{\sigma}_x \hat{X})^k/k!$. This approximation, however, is limited in the faithful reproduction of a strong RI, and a higher-order approximation is thereby required.  For example, the induced operation on the CV mode by a projective measurement on the DV mode after $\hat{U}^{(1)}$    exhibits at most a linear operation $x_0+ \hat{X}$ for a constant $x_0$, and therefore cannot emulate  higher orders of nonlinear operation  $\hat{O}_R$.  

\subsection{Second-order approximation}
\subsubsection{Two-mode squeezed vacuum ancilla}
We can access the second order approximation $\hat{U}^{(2)}(t)$ by simply applying an additional Gaussian squeezing on the oscillator as follows:
\begin{align}
&\hat{U}^{(2)}(t)= 1+i t \hat{\sigma}_x \hat{X}+\frac{(it)^2}{2}\hat{X}^2\approx(1+ it \hat{\sigma}_x \hat{X})e^{-t^2\hat{X}^2/2}.
\label{eq:expansion}
\end{align}
This factorization is possible due to the commutative property of all participating operators, and the involutority of the Pauli matrix $\hat{\sigma}_x^2=\hat{1}$ which simplifies the second-order approximation as a factorized form. 
 The squeezing operation $\exp[-t^2 \hat{X}^2/2]$ can be deterministically implemented \cite{MiwaPRL2014}, or can be supplied conditionally by letting a vacuum ancillary state interact with the target mode and  a vacuum detection on the ancilla is postselected ~\cite{fiurasekPRA2012Filter}. 
Alternatively, this squeezing can be merged with the pre-squeezing of the input state, as will be explained later.
If this $\hat{U}^{(2)}$ is applied to  $\ket{\psi}$ and $\ket{\phi_0}=c_+\ket{+}-c_-\ket{-}$ represented in the eigenbasis of $\hat{\sigma}_x$, an output state is obtained as
\begin{align}
&\hat{U}^{(2)}\ket{\phi_0}\ket{\psi}\approx c_+\ket{+}(1+it \hat{X}-\frac{t^2}{2} \hat{X}^2)\ket{\psi}\nonumber\\
&+c_-\ket{-}(1-it \hat{X}-\frac{t^2}{2} \hat{X}^2)\ket{\psi}\nonumber\\
&\approx e^{-\frac{t^2}{2} \hat{X}^2}\{(c_+\ket{+}+c_-\ket{-})\ket{\psi}+it (c_+\ket{+}-c_-\ket{-})\hat{X}\ket{\psi}\}\nonumber\\
&\equiv e^{-\frac{t^2}{2} \hat{X}^2}\left(\ket{\phi_0}\ket{\psi}+it \ket{\phi_1}\hat{X}\ket{\psi}\right).
\label{eq:altanative}
\end{align}
 where $\ket{\phi_1}=\hat{\sigma}_x \ket{\phi_0}=c_+\ket{+}-c_-\ket{-}$ and  $\ket{\pm}_d=2^{-1/2}(\ket{0}_d\pm\ket{1}_d)$ for optical qubits. Our principal case of interest on the last line is an equally weighted state $c_+=\pm c_-=2^{-1/2}$, which will be used later for verification of the Rabi dynamics. In this special case, the two qubit states $\ket{\phi_{0}}$ and $\ket{\phi_{1}}$ are orthogonal. For the proof-of-principle test, we may associate these qubit states  to the vacuum $\ket{\phi_{0}}=\ket{0}_d$ and single photon state $\ket{\phi_{1}}=\ket{1}_d$, or alternatively the vacuum $\ket{\phi_{0}}=\ket{1}_d$ and single-photon state $\ket{\phi_{1}}=\ket{0}_d$ for different setups to be explained. 
Therefore, $\hat{U}^{(2)}(t)$ can be achieved by a conditional application of an operation $i t \hat{X}$ controlled by the optical qubit state.

\begin{figure*}[th]
\includegraphics[width=500px]{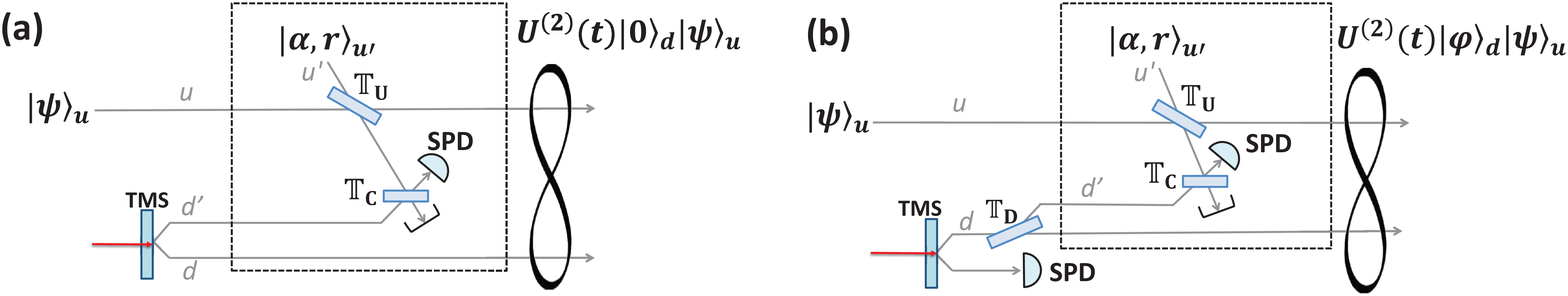}
\caption{All-optical setups for the implementation of an approximate RI $\hat{U}^{(2)}(t)$ on weak states of light.  
In both setups, an arbitrary state $\ket{\psi}_u$  in the upper arm experiences a second-order phase-dependent displacement  controlled by the state of a qubit in the lower arm.  The ancilla in auxiliary mode $u'$  chosen as a vacuum $\ket{\alpha,r}_{u'}=\ket{0}_{u'}$ mediates the two arms via a beam splitter U and C. All the beam splitters are denoted by their transmission coefficients $\mathbb{T}_i$. %, and generates the second order implementation of RI $\hat{U}^{(2)}$.   
A beam splitter C and the central single photon on-off detector (SPD) erases the which-path information of the ancillas and triggers a successful RI  by entangling the direction of approximate displacement in the phase space of $u$ and the qubit state in $d$.
The qubit state corresponding to an input photonic qubit $c_+\ket{+}_d+c_-\ket{-}_d$ in the lower arm is  from (a) a two-mode squeezed vacuum (TMSV) ($\ket{\mathrm{TMSV}}_{d,d'}\approx\ket{0}_{d}\ket{0}_{d'}+\lambda \ket{1}_{d}\ket{1}_{d'}$) with an adjustable weight $\lambda \ll 1$, or (b) a split single photon state.  The scheme in (a) is limited by the weak strength $\lambda$ constraint and  is mostly faithful for the simulation of states in the vicinity of $\ket{0}_d$. In comparison, the scheme with a split single photon in (b) allows to access all cases of input qubit states faithfully $\ket{\phi}_d$ by swapping the basis of TMSV and adjusting the coefficients of beam splitter D.  %, whose beam splitting ratio is $\cot^2 T_i$.
 The target strength of the achieved RI  can be chosen freely, and the coefficients $c_+,c_-$ are decided by controlling $\lambda$ and $\mathbb{T}_i$'s. % as in (\ref{eq:lambda},\ref{eq:lambdab}). 
The parameters for the TMSV in the lower arm and the strength of the beam splitter U are chosen in the entire manuscript as  $\lambda=0.01$, $\kappa=0.1$ of (\ref{eq:transform}). 
   The photon loss was added at the outputs of the mode $u$  to check the stability of our schemes. 
  }
\label{scheme}
\end{figure*}

In Fig.~\ref{scheme}, we propose setups that can implement the all-optical  entangling properties of  $\hat{U}^{(2)}(t)$ for weak states of light.
These setups are composed of  two arms of the process modes $u,d$, and two ancillary modes $u',d'$ which will be transmitted to a detection module at the center. 
In the upper arm, a displaced squeezed vacuum state $\ket{\alpha,r}_{u'}=D_{u'}[\alpha]S_{u'}[r]\ket{0}_{u'}$ interacts with an arbitrary CV input state $\ket{\psi}_u$  through a beam splitter U,  where a displacement is written as  $D[\alpha]=\exp[\alpha \hat{a}^\dagger-\alpha^* \hat{a}]$ and a squeezing operator is written as $S[r]=\exp[-\tfrac{r}{2} \hat{a}^{\dagger2}+\tfrac{r}{2}\hat{a}^2]$.   We assume the coherent amplitude $\alpha$ and squeezing parameter $r$ are real, as the general case of non-real $\alpha$ and $r$ does not affects the overall physics other than the success rate of the scheme.  
  Into the lower arm, a weak two-mode squeezed vacuum (TMSV) is injected approximating a DV entangled state $\ket{\mathrm{TMSV}}_{d,d'}\approx\ket{0}_d\ket{0}_{d'}+\lambda\ket{1}_d\ket{1}_{d'}$.   One mode $d$ represents the processing mode in DV, while the other auxiliary mode $d'$ plays the role of an ancillary mode.  To confine a weak TMSV into a qubit space, the coefficient is limited to $\lambda\ll 1$.   %for the proof-of-principle test.  
   
 A beam splitter with index $j$  mixes two optical modes $1$ and $2$ denoted by an interaction operator  $\exp[iT_j (\hat{a}^\dagger_1 \hat{a}_{2}+\hat{a}_1 \hat{a}^\dagger_{2})]$ with strength $T_j$. This beam splitter has a transmittance $\mathbb{T}_j=\cos T_j$ and a reflectance $\mathbb{R}_j=\sin T_j$, thus a beam splitting ratio $\cot^2 T_j$.   We first note that a beam splitter U can be converted  by applying suitable single mode pre- and post-squeezing $S_1[\pm r_\mathrm{tr}]$ on one of the modes into another useful form as %=\exp[-\tfrac{r_\mathrm{tr}}{2} \hat{a}^{\dagger2}+\tfrac{r_\mathrm{tr}}{2}\hat{a}^2]
\begin{align}
S_1[r_\mathrm{tr}]\exp[iT_\mathrm{U}(\hat{a}_1 \hat{a}_2^\dagger+\hat{a}_1^\dagger \hat{a}_2)]S_1[-r_\mathrm{tr}]\approx\exp[i\kappa \hat{X}_1 \hat{X}_2]
\label{eq:transform}
\end{align}
 with $\kappa=2 T_\mathrm{U}\cosh r_\mathrm{tr}$
 when $r_\mathrm{tr}\gg 0$. 
 For an experimental proof-of-principle test, the fixed pre-squeezing can be included in the state preparation, while the fixed post-squeezing can be performed on the data from quantum tomography of the output states. We note that this \textit{numerical squeezing} is possible as the Gaussian pre- and post-squeezing preserves quantum non-Gaussian aspects of the RI.

An avalanche single-photon detector (SPD) is the core source of the nonlinear effect in our scheme. On-off detection events of a single SPD are mathematically described by POVM elements $\{\hat{1}_{u'}-\ket{0}_{u'}\bra{0}, \ket{0}_{u'}\bra{0}\}$ in the ancillary mode $u'$. When the detected states  are weak in their intensity and the many-photon components are negligible, this POVM set can be approximated as $\{\ket{0}_{u'}\bra{0}, \ket{1}_{u'}\bra{1}\}$. Furthermore, if even single photon component is small, not performing any measurement and tracing out the state can  approximate  the detection of vacuum with an SPD as $\ket{0}_{u'}\bra{0}\approx \hat{1}$. In all the following examples, these weak intensity conditions are satisfied and the approximations about SPDs will be utilized.

   After a converted beam splitter  U in (\ref{eq:transform}) between  $\ket{\psi}_u$ and  a  weak displaced squeezed vacuum $\ket{\alpha,r}_{u'}$ with $\alpha, r\ll 1$,       an operation on mode $u$ is  induced when a SPD placed on ancillary mode $u'$ registers a single photon as
\begin{align}\label{opsingle}
&\hat{O}_1
=\frac{\sqrt{2} \left(\sqrt{2} e^{2r} \alpha+i \kappa\hat{X}_u\right) }{(e^{2r}+1)^{3/2}}
e^{-\frac{2 e^{2r} \alpha ^2}{2 e^{2r}+2}+\frac{2 i \sqrt{2} e^{2r}
   \kappa \alpha \hat{X}_u }{2 e^{2r}+2}-\frac{\kappa^2 \hat{X}_u^2}{2 e^{2r}+2}}
\end{align}
from a simple algebra.
 Here, a first-order polynomial in $\hat{X}$ (X-gate) \cite{ParkPRA2014}  $\sqrt{2} e^{2r} \alpha+i \kappa\hat{X}_u$ is contained  with a redundant Gaussian function of the quadrature operator $X_u$ up to a constant together. This applied X gate  is programmable by the parameters of the Gaussian state $|\alpha,r\rangle_{u'}$, with a flexibility in the value of $\alpha$ and $r$.  The simplest expansion in (\ref{eq:altanative}) can be achieved by simply setting $\alpha=0$, and $r=0$, while optimization of $\alpha$ and $r$ helps to reach a simulated interaction closer to an ideal Rabi gate  at a higher success probability. 
On the other hand, the detection of a vacuum at the SPD (approximated by completely tracing out) applies a Gaussian operation of $\hat{X}_u$ on the mode $u$ in the following form
\begin{align}
&\hat{O}_0
=\frac{1}{\sqrt{e^{2r}+1}}e^{-\frac{2 e^{2r} \alpha ^2}{2 e^{2r}+2}+\frac{2 i \sqrt{2} e^{2r}
   \kappa \alpha \hat{X}_u }{2 e^{2r}+2}-\frac{\kappa^2 \hat{X}_u^2}{2 e^{2r}+2}}.
\end{align}

In order to  simulate an operator $it \hat{X}$ and an identity operation $\hat{1}$ more faithfully  by $\hat{O}_1$ and $\hat{O}_0$ respectively, a correction for the redundant Gaussian operation needs to be included. The common factor $e^{-\frac{2 e^{2r} \alpha ^2}{2 e^{2r}+2}}/\sqrt{e^{2r}+1}$ in $\hat{O}_1$ and $\hat{O}_0$ influences only the overall success probability, and does not affect the physical properties of the output state if omitted.  The redundant displacement $e^{\frac{2 i \sqrt{2} e^{2r}
   \kappa \alpha \hat{X}_u }{2 e^{2r}+2}}$ can be also simply canceled by an additional inverse displacement operation applied  either before or after the interaction. The common redundant  Gaussian squeezing $e^{-\frac{\kappa^2 \hat{X}_u^2}{2 e^{2r}+2}}$ can be compensated by adding an additional anti-squeezing operator $S[r_\mathrm{corr}]$ with a squeezing parameter $r_\mathrm{corr}=-\log[\frac{\kappa^2}{e^{2r}+1}+1]/2$ either before or after the beam splitter U. Instead, it can be actively exploited as the source of squeezing required for the second order expansion $\hat{U}^{(2)}(t)$. 
These corrections can again be performed numerically on tomogram of the output Wigner function for the proof-of-principle experiments. 
   Therefore, the total operations conditionally applied together with the corrections are reduced into simpler forms
   \begin{align}
   \hat{O}_1=\frac{ i \kappa\hat{X}_u }{\sqrt{2}}, ~~~\hat{O}_0=\hat{1}.
   \label{eq:ops}
   \end{align}
  %  Again, free parameters $\kappa$ can be exploited for the optimal implementation of the target Rabi operation with different success probabilities and fidelities.

Into the other arm of two-mode $d$ and $d'$, a state with correlation in photon number is injected, such as TMSV in Fig.~\ref{scheme} (a). The detection module at the center is composed of a beam splitter C and one SPD placed at one of the output ports. The beam splitter erases the which-path information of the two ancillary modes $u'$ and $d'$.  We  post-select the detection event of a photon and the other mode being traced out as in \cite{MorinNat2014}. The tracing out is nearly equivalent to the vacuum projection, since a photon detection nearly heralds a non-existence of the photon in the other mode, i.e. if the detected photons are from $d'$,  the ancilla from $u'$ will contain no photons, and vice versa. 
   Due to the initial correlation in photon number   in TMSV, this post-selected detection outcome is also correlated with the binary photon number $\{\ket{0}_d,\ket{1}_d\}$ in the lower arm. The total effect of the setup is summarized as a weak approximate X-gate  $\hat{O}_1$ applied to $\ket{\psi}_u$ for a vacuum state  $\ket{0}_d$, and an identity operation $\hat{O}_0$ for a single photon state $\ket{1}_d$. Thus the second order approximation of Rabi gate $\hat{U}^{(2)}$ with the final state $\hat{O}_1\ket{\psi}_u\ket{0}_d+\lambda\hat{O}_0\ket{\psi}_u\ket{1}_d$ is achieved by the superposition of these terms if the beam splitter C is balanced. 
    A more general state $\mathbb{T}_\mathrm{C}\hat{O}_1\ket{\psi}_u\ket{0}_d+\mathbb{R}_\mathrm{C}\lambda\hat{O}_0\ket{\psi}_u\ket{1}_d$ can be achieved by an unbalanced central beam splitter C achieving a general projection $\mathcal{N}(\mathbb{T}_\mathrm{C}\bra{10}_{u'd'}+\mathbb{R}_\mathrm{C}\bra{01}_{u'd'})$ with a normalization factor $\mathcal{N}$ reflecting the conditional nature of the scheme. The simulated state $\ket{\Psi}=\exp[i t \sigma_x \hat{X}_u] \ket{\psi}_u(c_1\ket{1}_d+c_0\ket{0}_d)$ is then identified with coefficients
 \begin{align}
 c_1=\mathcal{N} \mathbb{R}_\mathrm{C} \frac{\lambda}{\sqrt{2}}, ~~~c_0=\mathcal{N} \mathbb{T}_\mathrm{C} \frac{\kappa}{2t }
 \label{eq:lambda}
 \end{align} 
 for the target strength $t$ . We again note that there are practical limitation in the parameters such as $\mathbb{T}_\mathrm{C}$, $\lambda$, and $\kappa$. For example,  higher Fock elements of the TMSV contribute unwanted terms, or the success probability of the scheme may be too low if $\mathbb{T}_\mathrm{C}$ is  too large or small. 
 
\subsubsection{Single-photon ancilla}
 %Other such example is . 
 Alternatively as in  Fig. \ref{scheme} (b), a single photon entangled state (or a dual-rail qubit) may be injected into the lower arm for the physically swapped entangled states from TMSV's case. A high-quality single photon state can be generated from  weak two-mode parametric squeezing  processes heralded by a SPD in one of the modes~\cite{LeJeannic2017,Morin2012Singlephoton}. 
 After a beam splitter D, this single photon is split into an entangled state $\mathbb{T}_\mathrm{D}\ket{1}_d\ket{0}_{d'}+\mathbb{R}_\mathrm{D}\ket{0}_d\ket{1}_{d'}$. % with beam splitter transmission coefficient $T_\mathrm{D}$.  
 Equivalently as in TMSV's case, an entanglement with an exchanged DV basis associated with operators $\hat{O}_1, \hat{O}_0$ will be generated as $\mathcal{N}' (\mathbb{T}_\mathrm{D}\hat{O}_1\ket{\psi}_u\ket{1}_d+\mathbb{R}_\mathrm{D}\hat{O}_0\ket{\psi}_u\ket{0}_d)$  by this setup. The target coefficients in  $\ket{\Psi}=\exp[i t \sigma_x \hat{X}_u] \ket{\psi}_u(c_1\ket{1}_d+c_0\ket{0}_d)$ are given as
 \begin{align}
c_1=\mathcal{N}' \mathbb{T}_\mathrm{D} \frac{\kappa}{2t }, ~~~ c_0=\mathcal{N}'  \frac{\mathbb{R}_\mathrm{D}}{\sqrt{2}}.
 \label{eq:lambdab}
 \end{align} 
We can also use the unbalanced beam splitter C as in the case of TMSV. 
 These two forms of ancillary entangled states, TMSV and a split single photon at the lower arm, cover the  RI on  full cases of the arbitrary DV states.  On the other hand, the split single photon ancilla enables to achieve the simulation beyond that with TMSV as  $\mathbb{T}_\mathrm{D}$ is not limited in contrast to $\lambda$, and the quality of the implementation is improved due to the suppressed high-photon number components.   In short, the achieved state can be simply written as 
\begin{align}
 &\ket{\Psi}= e^{-\frac{t^2}{2} \hat{X}_u^2}\big(c_+\ket{+}_d(1+i t \hat{X}_u)\ket{\psi}\nonumber\\
 &+c_-\ket{-}_d(1-i t \hat{X}_u)\ket{\psi}_u\big)
 \label{eq:u2}
 \end{align}
   for fully controllable values of the implemented RI strength $t$  for the simulation of the various weighted qubit state of $c_+\ket{+}_d+c_- \ket{-}_d$ with $c_{\pm}\in \mathbb{C}$ without loss of generality. The hybrid nature of RI in these optical setups is created from the continuous nature of the ancillas and the discrete nature of the photon detector. 
% We also note that we can utilize an unknown dual-rail or photon polarization qubit $a\ket{H}_d+b\ket{V}_d=c_1\ket{1}_d\ket{0}_{d'}+c_0\ket{0}_d\ket{1}_{d'}$ in an equivalent way. %as in Fig. \ref{scheme} (d).

\subsection{Third-order approximation}

For the simulation of a higher-strength Rabi gate  and its nonlinear effects, an approximation at a higher-order expansion of the RI $\hat{U}^{(j\ge 3 )}$  needs to be realized.  The schemes in Fig.~\ref{scheme}, however, are not immediately repeatable to increase the net strength, as one mode of the bipartite DV input encodings is consumed by the detection, and the output state has a single-rail qubit encoding with one less mode. Therefore an additional converter or an entangler is required to convert it back to the dual-rail-like encoding. An existing deterministic scheme is based on a hybrid teleportation through a two photon entanglement or GHZ-like entanglement generated from a parametric nonlinearity~\cite{fiurasekPRA2017}.

 Feasible approaches toward accessing a high-strength RI encompass using  more complex ancillas, or a more complex setup.
  The first approach is inspired by the fact that a coherent state ancilla can induce a displacement operation on the target mode by letting the modes $d'$ and $u$ interact through a beam splitter U  and the vacuum detection on mode $d'$ afterward as $_{d'}\bra{0}\exp[i\kappa\hat{X}_u\hat{X}_{d'}]\ket{\alpha}_{d'}=e^{\frac{1}{4} \left(-2 \alpha ^2-\kappa^2 \hat{X}_{d'}^2+2 i \sqrt{2} \alpha  \kappa \hat{X}_{d'}\right)}\approx e^{i \alpha  \kappa \hat{X}_u/\sqrt{2}}$, again when the Gaussian squeezing term is ignored due to small $\kappa$ or an application of anti-squeezing.  
  Therefore, an entangled ancilla of the form $\ket{0}_{d}\ket{\alpha}_{d'}+\ket{1}_{d}\ket{-\alpha}_{d'}$ is directly translated  into the ideal output form $\ket{1}_d\exp[i\kappa \hat{X}_u]\ket{\psi}_u+\ket{0}_d\exp[-i\kappa \hat{X}_u]\ket{\psi}_u$ (or the ancilla $\ket{+}_{d}\ket{\alpha}_{d'}+\ket{-}_{d}\ket{-\alpha}_{d'}$   into $\ket{+}_d\exp[i\kappa \hat{X}_u]\ket{\psi}_u+\ket{-}_d\exp[-i\kappa \hat{X}_u]\ket{\psi}_u$). 
  There exists experimental schemes to generate an approximation of these hybrid ancillas  utilizing the closeness of  a superposition of coherent states  $\ket{\alpha_\pm}=N_\pm(\ket{\alpha}\pm\ket{-\alpha})$ to a photon-added or subtracted squeezed vacuum state $\hat{a}^\dagger\ket{\gamma}$, $\hat{a}\ket{\gamma}$ or squeezed single photon $S[\gamma]\ket{1}$ with a certain squeezing parameter $\gamma$~\cite{AndersenPRA2013,MorinNat2014,JeongNat2014}.
   This analogy can be improved further by subtracting or adding more photons or using higher Fock states as  $S[\gamma]\ket{n=even}\approx \ket{\alpha_+}$ and $S[\gamma]\ket{n=odd}\approx \ket{\alpha_-}$, or $\hat{a}^{n=even}\ket{\gamma}\approx \ket{\alpha_+}$ and $\hat{a}^{n=odd}\ket{\gamma}\approx \ket{\alpha_-}$.  However, there exist two limitations: the vacuum detection is not heralded, and the approximation based on photon subtracted squeezed states need multiple photon subtractions is demanding experimentally.

\begin{figure*}[th]
\includegraphics[width=500px]{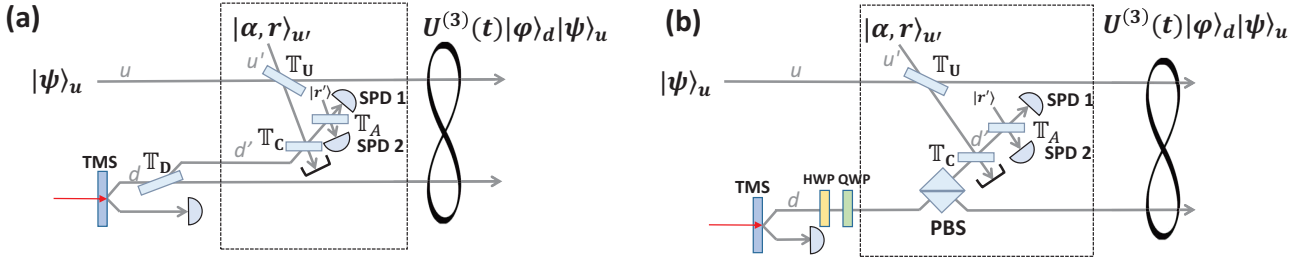}
\caption{
  A higher order approximation $\hat{U}^{(3)}$ can be achieved equivalently for (a) a dual-rail encoding and (b) a polarization qubit encoding by inserting an auxiliary squeezed state $|r'\rangle$ and a highly transmissive beam splitter A in front of the detectors.  The parameters for the beam splitter U, C, and D are the same as before. 
 The auxiliary squeezing parameter was chosen as $r'=-1.04$ corresponding to $6.02$ dB.  In both setups, the photon loss was considered on the mode $u$ outputs.  
  }
\label{schemeU3}
\end{figure*}

Alternatively, we are resorting to the second approach utilizing a more complex setup to achieve the third order approximation $\hat{U}^{(3)}= 1+i t \hat{\sigma}_x \hat{X}-\frac{t^2}{2}\hat{X}^2-i\frac{t^3}{6}\hat{\sigma}_x\hat{X}^3$. We first note that additional squeezing in the form of $\exp[-\xi_\pm \hat{X}^2]$  between the even-order terms $\cos[t\hat{X}]\approx1-\frac{t^2}{2}\hat{X}^2\approx \exp[-\frac{t^2}{2}\hat{X}^2]$ and odd-order terms $\sin[t\hat{X}]\approx t\hat{X}-\frac{(t\hat{X})^3}{6}=t\hat{X}(1-\frac{(t\hat{X})^2}{6})\approx t\hat{X}\exp[-\frac{(t\hat{X})^2}{6}]$  are different in the strengths $\xi_+=t^2/2$ and $\xi_-=t^2/6$. Therefore, a squeezing controlled by an ancillary qubit state is essential in order to achieve the $\hat{U}^{(3)}$. 
The approximation of $\hat{U}^{(3)}$ containing the required controlled squeezing can be achieved by adding a measurement-induced online squeezing in front of the central SPD as in Fig.~\ref{schemeU3}. A squeezed state $\ket{r'}_a$ in the additional mode $a$ interacts with ancillary mode $d'$ through an additional beam splitter A, and second detector (SPD 2) registers photons on this mode.  In a high transmitivity limit $\mathbb{T}_A\rightarrow 1$  of the  beam splitter A, the single photon detection  is suppressed due to the photon number parity selection rule for the squeezed state. Therefore, the photon detection is mostly from the two-photon detection which  applies an effective squeezing $\exp[-\zeta \hat{X}_{u'}^2]$ with $\zeta=-\frac{\left(4 e^{2 r' }+5\right) \kappa'^2}{2 \left(e^{2 r' }+1\right)}$ where $\kappa'$ is the strength of the transformed beam splitter A as in (\ref{eq:transform}).  This squeezing  is transferred  to the mode $u$ selectively only when the detected photon at SPD 1 is from mode $u'$, while does nothing if it is from the mode $d'$.   The theoretical description of the net operations applied to the mode $u$ in Eq.~(\ref{eq:ops}) is changed  as:
 \begin{align}
 &  \hat{O}_1^{(3)}=_{u'}\bra{1}\exp[-\zeta \hat{X}_{u'}^2]\exp[i\kappa \hat{X}_{u'}\hat{X}_{u}]\ket{0}_{u'}=\frac{i \kappa  \hat{X}_u e^{-\frac{\kappa ^2 \hat{X}_u^2}{4 \zeta +4}}}{\sqrt{2} (\zeta +1)^{3/2}}\nonumber\\
 &  \hat{O}_0^{(3)}=_{u'}\bra{0}\exp[i\kappa \hat{X}_{u'}\hat{X}_{u}]\ket{0}_{u'}=e^{-\frac{1}{4} \kappa^2 \hat{X}_{u}^2}.
   \end{align}
 Here, note that the squeezings included in these two operations are different as required. The final forms of the output states are therefore given by 
\begin{align}
&\mathbb{T}_\mathrm{C}\ket{1}_d \hat{O}_0^{(3)} \ket{\psi}_u+\mathbb{R}_\mathrm{C}\lambda\ket{0}_d \hat{O}_1^{(3)} \ket{\psi}_u, \nonumber\\
&\mathbb{T}_\mathrm{D}\ket{1}_d\hat{O}_1^{(3)}\ket{\psi}_u+\mathbb{R}_\mathrm{D}\ket{0}_d\hat{O}_0^{(3)}\ket{\psi}_u 
\label{eq:U3}
\end{align}
depending on whether a TMSV with the unbalanced detection module or a single photon ancilla is used respectively. The parameters for these achieved states to faithfully reproduce the ideal form $\ket{1}_d\cos[t \hat{X}_u]\ket{\psi}_u+\ket{0}_di \sin[t \hat{X}_u]\ket{\psi}_u \approx \ket{1}_d\exp[-\frac{t^2}{2}\hat{X}_u^2]\ket{\psi}_u+\ket{0}_di t\hat{X}_u\exp[-\frac{(t\hat{X}_u)^2}{6}]\ket{\psi}_u$  are $\kappa=\sqrt{2}t$, $\zeta=2$ and $\lambda\mathbb{T}_\mathrm{C}/\mathbb{R}_\mathrm{C}=3\sqrt{3}$. 
This strength of squeezing $\zeta$ is again achieved by the squeezing parameter $r'=-1.04$, corresponding to $6.02$ dB. We remind again that additional pre- or post-squeezing allows a wider range of parameters of the setup  as for the second order approximation.

\section{Verification of hybrid dynamics beyond RWA}
\label{secIII}

In this section, in order to witness that the achieved approximate interactions fit an ideal RI and possess the nonlinear nature beyond the RWA, we show that  the output states escape from the confined energy subspace imposed by the JC interaction~\cite{JC,JC2,JC3} for the exemplary input states.  For a direct comparison between  the realized state $\rho_\mathrm{re}$  and the ideal target states $\rho_\mathrm{id}$ produced either by a JC interaction or a RI, we calculate the fidelity $F=\mathrm{Tr}[\sqrt{\sqrt{\rho_\mathrm{id}}\rho_\mathrm{re}\sqrt{\rho_\mathrm{id}}}]^2$ as the closeness measure. In addition, the output states from RI and JC interaction reside in the bipartite system, and  their nature can be precisely characterized by the amount of entanglement for the differentiation of the two evolutions.  As a measure of feasibility of the experimental realization, we analyze the  success probability of our protocols.

\subsection{Entanglement vs. energy}
A quantum non-demolition dynamics of $\hat{X}_{\theta}$ and $\hat{\sigma}_i$ by a controlled displacement of a RI is significantly reduced or destroyed by the RWA, under which the RI collapses to the JC interaction $\hat{U}_\mathrm{JC}(\tau)=\exp[i\tau (\hat{\sigma}_+\hat{a}+\hat{\sigma}_-\hat{a}^{\dagger})]$. As a result, the JC interaction is missing the control of a conjugate variable $\hat{P}_{\theta}$ by the state of the qubit system. 
Let us start by briefly describing the effect of   $\hat{U}_\mathrm{JC}(\tau)$ and $\hat{U}_\mathrm{Rabi}(t)$ on arbitrary input states $\ket{\psi}_u$. The eigenstates of the JC interaction Hamiltonian $H_\mathrm{JC}=\hat{\sigma}_+ \hat{a}_u+\hat{\sigma}_- \hat{a}^\dagger_u$ are given as a dressed form $\ket{n}_\pm=2^{-1/2}(\ket{n}_u\ket{e}_d\pm\ket{n+1}_u\ket{g}_d)$ with the eigenvalues $\pm\sqrt{n+1}$  for any integer $n$. With the substitution of the bases as $\{\ket{g},\ket{e}\}\rightarrow \{\ket{0}_d,\ket{1}_d\}$, a unitary evolution by the JC interaction can be described with local operators as 
\begin{align}
&\hat{U}_\mathrm{JC}(\tau)=\cos[\tau\sqrt{\hat{n}_u+1}]\otimes\ket{1}_d\bra{1}+\cos[\tau\sqrt{\hat{n}_u}]\otimes\ket{0}_d\bra{0}\nonumber\\
&+i\frac{\sin[\tau\sqrt{\hat{n}_u+1}]}{\sqrt{\hat{n}_u+1}}\hat{a}_u\otimes\ket{1}_d\bra{0}+i\frac{\sin[\tau\sqrt{\hat{n}_u}]}{\sqrt{\hat{n}_u}}\hat{a}^\dagger_u\otimes\ket{0}_d\bra{1}\label{eq:JCloc}
\end{align}
for simpler calculations. For the two-mode input state $\ket{\psi_\mathrm{in}}=\ket{+}_d\ket{\psi}_u$ made of an arbitrary CV state  in Fock representation $\ket{\psi}_u=\sum_n h_n\ket{n}_u$ and a balanced DV state  $\ket{+}_d$, we obtain the output state decomposed into the eigenstates of the  JC interaction as
\begin{align}
&\ket{\Psi_{JC}}=\hat{U}_\mathrm{JC}(\tau)\ket{\psi_\mathrm{in}}=\hat{U}_\mathrm{JC}(\tau) \sum_n h_n\ket{n}_u\frac{\ket{1}_d+\ket{0}_d}{\sqrt{2}}\nonumber\\
&=\frac{h_0}{\sqrt{2}}\ket{0}_u\ket{0}_d\nonumber\\
&+\sum_n \frac{(h_{n+1}+h_n)e^{i\tau\sqrt{n+1}}(\ket{n}_u\ket{1}_d+\ket{n+1}_u\ket{0}_d)}{2\sqrt{2}}\nonumber\\
&+\sum_n\frac{(-h_{n+1}+h_n)e^{-i\tau\sqrt{n+1}}(\ket{n}_u\ket{1}_d-\ket{n+1}_u\ket{0}_d)}{2\sqrt{2}}.\label{eq:arbJC}
\end{align}
In this form, the complex rotation of the coefficients in each energy subspace is evident. This energy confinement is the characteristic of a  JC interaction, which holds for  cases when the field frequency matches the qubit excitation frequency. 
The evolution of the simplest case of vacuum input state in CV mode $\ket{\psi_\mathrm{in}}=\ket{+}_d\ket{0}_u=\frac{\ket{1}_d+\ket{0}_d}{\sqrt{2}}\ket{0}_u$ is  described as
\begin{align}
&\ket{\psi_{JC}}=\hat{U}_\mathrm{JC}(\tau)\frac{\ket{1}_d+\ket{0}_d}{\sqrt{2}}\ket{0}_u\nonumber\\
&=\frac{1}{\sqrt{2}}(\cos \tau\ket{1}_d\ket{0}_u+\sin \tau\ket{0}_d\ket{1}_u)+\frac{1}{\sqrt{2}}\ket{0}_d\ket{0}_u, \label{eq:JCout}
\end{align}
where the output state is confined in the energy subspace of a single and zero quantum.

To find a measure of the faithfulness of the simulation, we can use the fact that the output states are bipartite entangled states. 
The analysis on the entanglement reflects a nature of the states which cannot be quantified by the analysis on the local states separately. The negativity as a measure of entanglement \cite{neg,neg2} of a bipartite state density operator $\rho$ is calculated as $N[\rho]=\frac{\mathrm{Tr}[|\rho^\mathrm{PT}|]-1}{2}$, where $\rho^\mathrm{PT}$ is $\rho$ partially transposed and $\mathrm{Tr}[|\cdot|]$ stands for the trace norm. For a fair comparison of the entanglement generated by approximate processes,  we take test mixed state $\rho$  with the same initial  total number of quanta (or simply energy) $E[\rho]=\mathrm{Tr}[(\hat{n}_u+\hat{n}_d)\rho]$.   The negativity of the state in (\ref{eq:JCout}) is given by  a periodic function $N[\ket{\psi_{JC}}\bra{\psi_{JC}}]=|\sin 2\tau|/4$ and the maximal value $1/4$ at $\tau=\pi/4$, while the energy a fixed value $0.5$.
Another notable example is a vacuum input state in mode $u$ with an energy excited eigenstate in DV mode, i.e. $\ket{\psi'_\mathrm{in}}=\ket{1}_d\ket{0}_u$. In that case, the state evolves into $\ket{\psi'_{JC}}=\cos \tau\ket{1}_d\ket{0}_u+\sin \tau\ket{0}_d\ket{1}_u$, with the negativity $|\sin 2\tau|/2$ and the total energy stays equal to a single quantum. If the DV mode is in the ground state (or vacuum) $\ket{0}_d\ket{0}_u$,  the JC interaction does not affect the state and no entanglement is generated. In general, less entanglement tends to be generated by the vacuum state than the single photon state in qubit mode $d$  due to a lower number of quanta  for an arbitrary input state in mode $u$.

In comparison,  $\hat{U}_\mathrm{Rabi}(t)$ transforms an optical input state $\ket{\psi}_u$ and a qubit state $\ket{1}_d$ and $\ket{0}_d$ respectively into an entangled form as
\begin{align}
\ket{\Psi}_\mathrm{R}=\frac{1}{\sqrt{2}}\ket{+}_d\exp[it\hat{X}_u]\ket{\psi}_u\pm\frac{1}{\sqrt{2}}\ket{-}_d\exp[-it\hat{X}_u]\ket{\psi}_u.\label{eq:symRabi}
\end{align}
%We notice that the state in optical mode $u$ is displaced in the opposite directions in phase space depending on the entangled qubit basis states. 
 The negativity for vacuum input state $\ket{\psi}_u=\ket{0}_u$ can be calculated analytically in the qubit entanglement formalism given as $\frac{1}{2} \sqrt{1-e^{-2 t^2}}$, increasing for a larger $t$ asymptotically to a maximal entanglement due to the decreasing overlap of the displaced states in $u$. 
 The energy contained in this state is expressed as $ \left(t^2+1\pm e^{-t^2}\right)/2$. 
We note that when the DV mode $d$ is initially in the state $\ket{+}_d$ or $\ket{-}_d$, the input state in mode $u$ is simply displaced and  no entanglement is generated by both  the ideal and approximate RI.

Asymptotically, for sufficiently large $t$ both energy and negativity generated by the RI significantly break the limits of those by the JC interaction. 
  Therefore, the simultaneous escape from the energy constraint and the bound of  entanglement of the JC interaction can be a strong signature of the dynamics beyond the RWA. 
 The simplest evidence of non-RWA dynamics can be found in the case of the vacuum state  $\ket{0}_d\ket{0}_u$. In this case, as the JC interaction cannot generate any entanglement and energy remains minimal, even a slight shift of energy and amount of generated entanglement by any process implies its non-RWA dynamics. In order to show that our scheme reproduces the effects of general RI, a comparison in a broader set of input states can be performed using aforementioned quantifiers. 
 The  set of states chosen based on the easiness of theoretical description, experimental generation, and the Gaussian natures which makes the non-Gaussian nature of the dynamics more manifest, are coherent states $\ket{\beta}_u$, thermal states $\rho_\mathrm{th}[\bar{n}]=\left(\frac{\bar{n}}{\bar{n}+1}\right)^{\hat{n}_u}/(\bar{n}+1)$ with an average photon number $\bar{n}$  and phase-randomized states (PRC) $\rho_\mathrm{prc}[\beta]=\int_{\phi=0}^{2\pi} d\phi \ket{\beta e^{i\phi}}_u\bra{\beta e^{i\phi}}$.

\begin{figure*}[th]
\includegraphics[width=500px]{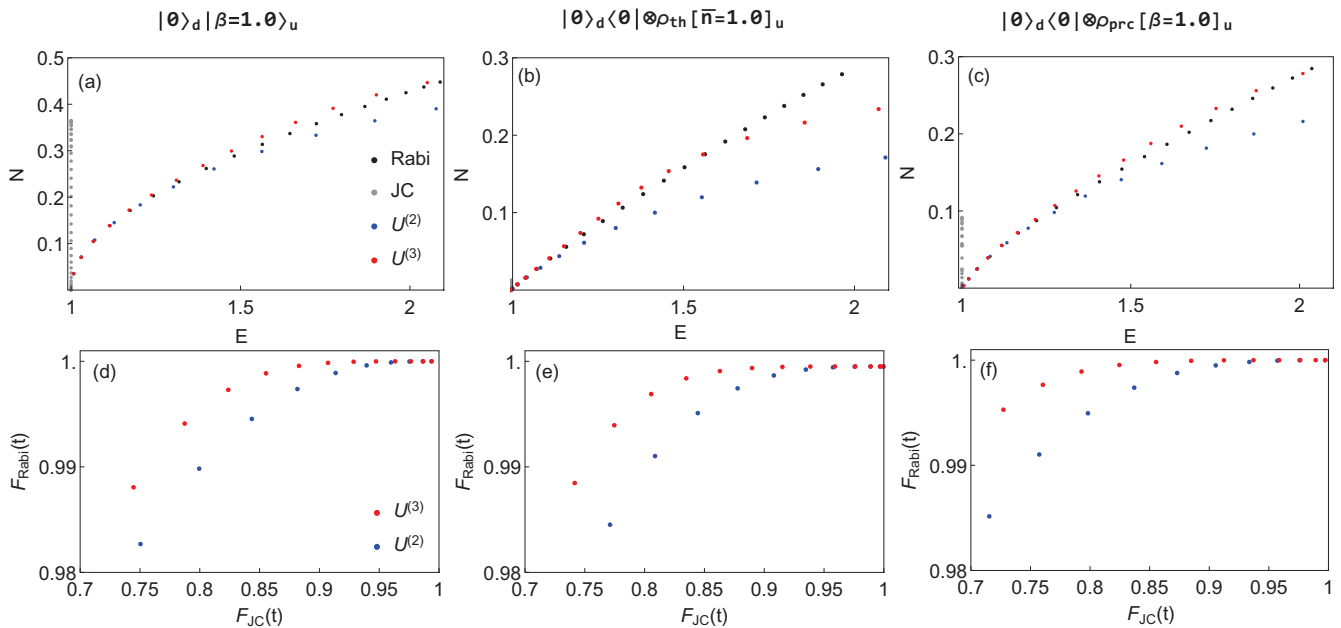}
\caption{Comparison of the  states generated by the ideal JC interaction, the ideal RI,  and by our setups in Fig. \ref{scheme} and \ref{schemeU3}. The input DV  state is a vacuum $\ket{0}_d$ in all cases.
   The input state in the upper mode $u$ is (a,d) a coherent state $\ket{\beta=1}_u$, (b,e)  a thermal state $\rho_\mathrm{th}[\bar{n}=1.0]$ and (c,f) a phase-randomized coherent state $\rho_\mathrm{prc}[\beta=1.0]$.  Dots were drawn at the strength interval of $\delta t=0.05$ starting from $t=\tau=0$. 
 (a-c)  In the diagrams of energy $E$ vs. negativity $N$, the total energy (the number of excitations) is conserved by the JC interaction, which shows a stark contrast to the ideal and approximate RIs which is  a clear signature of a dynamics beyond RWA. The behavior of the entanglement and energy by the ideal RI and our schemes agree well.  The third order approximation $\hat{U}^{(3)}$ has enhanced closeness the ideal RI. %For a thermal state in (b), the generated entanglement by JC interaction is negligible. 
   (d-f) The fidelity of output states generated by our schemes against ideal RI $F_\mathrm{Rabi}(t)$  is much higher than that against  the ideal JC interaction $F_\mathrm{JC}(t)$ for the varied strengths $t$. The fidelity of the generated state by  $\hat{U}^{(3)}$ against the ideal RI is improved even further. %The same experimental parameters are assumed as in Fig. \ref{scheme}.
     }
\label{fig:EvsN}
\end{figure*}

  In Fig.~\ref{fig:EvsN}, we compared the evolution of energy and entanglement of the states generated by the ideal JC interaction (\ref{eq:JCout}), the ideal RI (\ref{eq:symRabi}) and $\hat{U}^{(2,3)}$  by  our setups in Fig.~\ref{scheme} and \ref{schemeU3} at varied strengths. 
 We can first notice that the behavior of the curves made by the two ideal interactions on these states are significantly different, which can be summarized as follows. First, the total energy (the number of excitations) is conserved by the JC interaction regardless of input states. Second, the maximum amount of entanglement generated by the JC interaction does not reach the maximal value $1/2$, and is smaller than that by RI which reaches $1/2$ for any input state in the high-strength limit. 
These features can be used as the witnesses of a non-RWA nature of the implemented processes. We notice that the realistically generated states possess more energy and entanglement than the state generated by JC interaction at some strengths, and are fairly analogous to those of ideal RI in the general tendencies. 
In Fig.~\ref{fig:EvsN} (d-f), the fidelity of  the output state against  the state from ideal RI $F_\mathrm{Rabi}(t)=|\bra{\psi_\mathrm{in}}\hat{U}_\mathrm{Rabi}(t)\hat{U}^{(2)}(t)\ket{\psi_\mathrm{in}}|^2$    is significantly higher than against the state from ideal JC interaction $F_\mathrm{JC}(t)=|\bra{\psi_\mathrm{in}}\hat{U}_\mathrm{JC}(t)\hat{U}^{(2)}(t)\ket{\psi_\mathrm{in}}|^2$ in all cases. The generated states are much closer to the target states generated by the RI than those by JC interaction. If the third order approximation $\hat{U}^{(3)}$ is accessible, the ideal Rabi gate can be achieved with even a higher fidelity, and a higher agreement in the energy and the entanglement.

For the experimental proof-of-principle test, thermal states and phase-randomized coherent states  have an advantage as a witness  in comparison to coherent states,  as less entanglement is generated by the ideal JC interaction from them.
This is because a JC interaction in (\ref{eq:arbJC}) acts as a combination of a rotation-like transformation and a quanta exchange,  both of which have only weak effects on states whose density matrices are diagonal in Fock basis.  In Fig. \ref{fig:EvsN} (c), we notice that the generated amount of entanglement for PRC is  intermediate between coherent states and  thermal states.  We briefly add that states with higher photon numbers are less favored for the exhibition of non-RWA dynamics. In these cases, the generated entanglement by JC interactions gets larger for all classes of states although it cannot reach the maximal value of $1/2$. Moreover, a less amount of entanglement is generated by RI from such a state than a weak-intensity state due to the increased overlap between displaced states. We note that our scheme still can generate a larger entanglement than JC interaction even in this case. In all cases of input states and processes, the generated entanglement does not have any amount of Gaussian entanglement~\cite{Laurat2005Gaussian}, a necessary condition for hybrid entangled states \cite{KreisPRA2012}. 

\subsection{Success probability of the protocols}

For the complete description of our scheme, the success probability can be considered as a measure of the feasibility of an experimental implementation. For simplicity, we assume that  the resource states such as TMSV and single photon ancilla can be prepared offline and thus irrelevant to the success probability of the implementation. Also all the Gaussian operations such as squeezing and displacement operations are assumed to be applied deterministically. We note that the success probability of the unspecified experimental elements not mentioned in our scheme is not considered into account.
 
The total success probability of our schemes $P(t)$ at strength $t$ can be simply calculated theoretically by  taking the norm of the output states in (\ref{eq:u2}) and (\ref{eq:U3}) for the simplest case of $c_+=c_-$. Choosing a squeezed vacuum ancilla in mode $u'$ can increase the overall success probability, but can induce a reduction in the quality of the simulation. We therefore consider only a vacuum ancilla. We note that tracing out one of the modes in the central detection module reduces the total success probability by half.  

\begin{figure*}[th]
\includegraphics[width=500px]{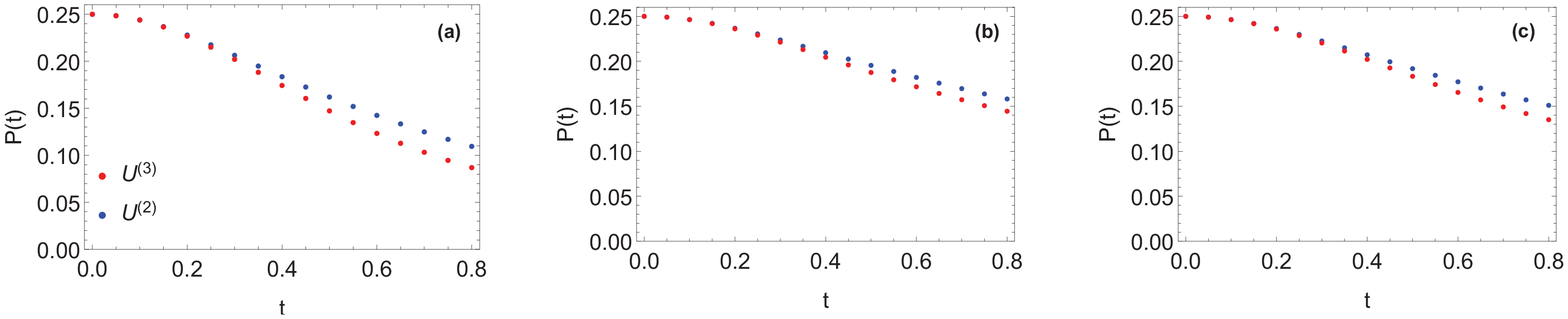}
\caption{
 Success probability of our scheme vs. strengths of RI  at an interval of $\delta t=0.05$ for (a) coherent states, (b) thermal states and (c) phase-randomized coherent states with average photon number $\bar{n}=1$. The success probability of the implementation of $\hat{U}^{(3)}$ is slightly lower than that of $\hat{U}^{(2)}$, and both decrease for a stronger $t$.
  }
\label{fig:Pvst}
\end{figure*}

In Fig. \ref{fig:Pvst}, success probability of our schemes for the test states are shown.  In general, an implementation at a higher strength $t$ is achieved with a decreased success probability from the implementation success probability $1/4$ of doing nothing at $t=0$. This success probability corresponds to the case where beam splitters are completely transmissive $\mathbb{T}_U=\mathbb{T}_A=1$, and thus the chance of the photon from the ancilla to be detected at mode $u'$ after the  beam splitter C is $1/4$. The success probability of the implementation of $\hat{U}^{(3)}$ is slightly lower than that of $\hat{U}^{(2)}$. We note that a lower success probability  is expected on a state with a higher photon number as well.

In the next section, we will investigate the local qualitative features beyond quantitative analysis to show how faithful  our simulation schemes are in the reproduction of quantumness in a realistic experiment.

\section{Remote steering of displacements under effects of photon loss}
\label{secIV}

A distinguishing feature of the RI resulting from its unique hybrid entanglement structure is the steering of displacement operations on the oscillator by a projective measurement onto conjugate bases of the output DV mode. 
From (\ref{eq:symRabi}), i.e. an output state  from an ideal RI, we obtain the following local states by projections:
\begin{align}
&~_d\braket{1}{\Psi}_\mathrm{R}=\frac{\exp[it\hat{X}_u]+\exp[-it\hat{X}_u]}{2}\ket{\psi}_u, \nonumber\\
&_d\braket{0}{\Psi}_\mathrm{R}=\frac{\exp[it\hat{X}_u]-\exp[-it\hat{X}_u]}{2}\ket{\psi}_u,
\end{align}
which shows the application of superpositions of two opposite displacement operations, the sign depending on the local detection outcomes $\{\bra{1}_d,\bra{0}_d\}$. We again note that this detection can be implemented approximately by a SPD detection for the weak lights. % A detector efficiency is still limited for the realistic superconducting detectors, but for which the dark counts are already negligible \cite{LeJeannic2017}. 
 In comparison,  if we project onto $_d\bra{\pm}$,  we get simply displaced states as $~_d\braket{\pm}{\Psi}_\mathrm{R}=\exp[\pm it\hat{X}_u]\ket{\psi}_u$ respectively. This type of detection can be achieved approximately by homodyne measurements with quadrature window selection~\cite{LeJeannicOptica2018}. 

To test the stability of our schemes, let us consider an effect of photon loss occurring at various places as an environmental effect, a major source of decoherence in optical experiments.
The photon loss on the input coherent states, PRC or thermal states in mode $u$ has a minor impact and and only changes the average photon numbers,   which can be simply re-adjusted in the experimental proof-of-principle test.  The loss acting on the entangled ancillas can be detrimental to the quality of the experiment, but the projective detection by SPDs can eliminate this effect for sufficiently weak states in principle, as the heralded arrival of photon guarantees the existence of a maximal single photon entanglement. We therefore assume that the loss on the input states can be completely undone. Similarly, the photon detection on mode $d$ for the steering by projection can eliminate the effect of loss in the lower arm. 
Only the loss in the output states  in the $u$ mode is uncontrollable, and we include $15\%$ of loss output mode as a realistic level of loss~\cite{CavaillesPRL2018HybridSteering}, modeled as a beam splitter interaction with a vacuum bath.

\begin{figure*}[th]
\includegraphics[width=500px]{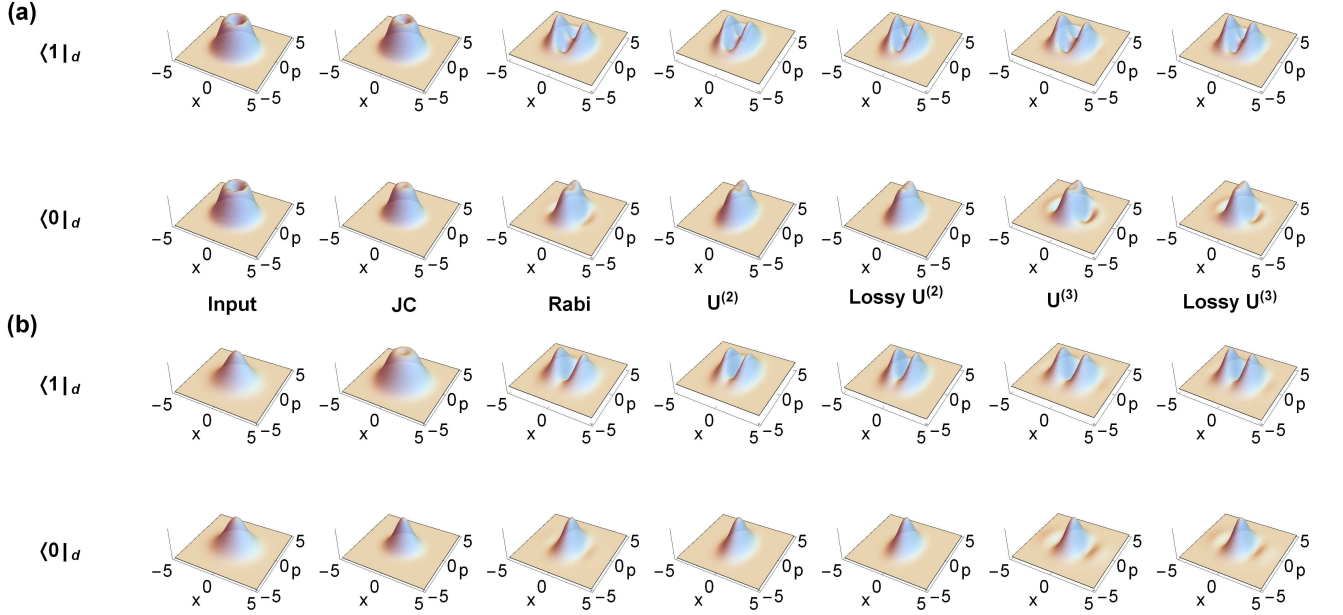}
\caption{Wigner functions of the input state, the output state after JC interaction $\hat{U}_\mathrm{JC}(\tau)$, RI $\hat{U}_\mathrm{Rabi}(t)$, the implemented interaction $\hat{U}^{(2)}$, advanced scheme of $\hat{U}^{(3)}$ with and without loss.  (a) An input phase-randomized coherent state and (b)  an input thermal state both with the same average photon number $\bar{n}=1$ after  (above) projection onto $\bra{1}_d$ and (below) onto $\bra{0}_d$.  The qubit input state is modeled as $\ket{0}_d$, while only  the two projection outcomes are exchanged for $\ket{1}_d$. 
The target strengths were set as $ \tau, t=0.7$ for the best visualization. Again a vacuum ancilla was assumed in mode $u'$, and the same experimental parameters are chosen as in Fig. \ref{scheme} and \ref{schemeU3}.
%with \B{$\mathbb{T}_\mathrm{C}=0.101$}, and $r'=-1.04$ for the third order approximation. 
 We notice that the steering by both projections on the output states from RI and engineered interactions have clear resemblance, and the agreement is improved as the order of approximation increases.  For both input states, the resemblance for the detection outcome $\bra{0}_d$ is in the generation of squeezing, while for the detection outcome $\bra{1}_d$, it is in the double peak structure. Third order approximation $\hat{U}^{(3)}$ improves the resemblance in the negative peaks. These negative dips are reduced but still visible at a $15\%$ photon loss applied on the output states.  All of these states are in a sharp contrast to the states resulting from the  JC interaction, which produces only a phase-invariant Wigner functions by both projections.
}
\label{fig:local}
\end{figure*}

Now we test the aforementioned steering features for thermal and phase-randomized coherent  states.  Both states are mixed, and exhibit a natural phase invariance and Gaussianity, and thereby the phase sensitivity and the non-Gaussian features in the Wigner function of the output states can be easily noticed. The initial photon number is arbitrarily fixed at $\bar{n}=1$ as an example. The initial qubit state was chosen as $\ket{0}_d$ as before.
 In Fig.~\ref{fig:local}, we show the output Wigner functions obtained by various processes after the qubit has been projected onto $\bra{1}_d$ or $\bra{0}_d$. In all cases, our setups generate similar features to the ideal RI. For both input states, a single squeezed peak and small negative regions are created in the case of the projection $\bra{0}_d$, and double peaks and a negative region are generated by the superposition of displacements in the case of the projection $\bra{1}_d$.   These features are in sharp contrast with the shapes of the Wigner functions of the  states generated by JC interaction steered by the same measurements. In contrast to the RI,  a rotation-invariant shape of the peak signifies a phase-invariant JC interaction. This similarity and contrast in the non-Gaussian features of Wigner functions holds true for a different $\bar{n}$.  In case of projection $\bra{0}_d$, a higher-order approximation $\hat{U}^{(3)}$ reproduces the negative Wigner functions of RI, implying that a higher-order approximation helps to get faithful effects on various states. The requirement of a larger number of subtracted photons (by more SPDs) motivates an advanced measurement setup in the router.  We note that the phase-sensitive negative dips, the key feature of the RI,  is still visible under a $15\%$ photon loss. This comparison exhibits that our scheme can generate  hybrid entanglement and reproduce the effects of RI beyond RWA from Gaussian excitations.

\section{Conclusion}
\label{secV}
In this work, we proposed  all-optical  schemes to implement the quantum RI its hybrid entangling effects in a heralded way. Our scheme is based on the realization of a controlled displacement on one mode (CV part) by the presence of photon in the other mode (DV part). The entangling effect arises from a joint photon detection on the single and two-mode squeezed ancillary states  interacting with these modes.  To verify that these setups can faithfully reproduce the core effects of a RI, we analyzed various aspects of the generated states: energy-entanglement relation, the fidelity against the state generated by the ideal RI, and a remote steering of displacement on CV part by a projective measurement on DV part.  The faithful tracking of energy-entanglement relation, high fidelity with ideal RI output states and the generation of phase-sensitive negative peaks in the Wigner functions of weak Gaussian states is a  conclusive witness of the experimental implementation of the RI beyond the RWA.   Our proposal reproduces the hybrid entanglement effects  of an ideal RI faithfully with a sufficient tolerance to expected experimental imperfections.  The proposal is feasible with current hybrid optical technology \cite{MorinNat2014,ULA2015Hybrid,JeongNat2014,RaTrepsARXIV2019non-Gaussianmultimode,MarekPRA2017HOM}. 
% We evidenced the hybrid entanglement generation and nonlinear natures of the  RI in the optical setups.  
This work opens the possibility of all-optical implementation of various nonlinear interactions which are currently available only at mechanical or microwave frequencies, and motivates advanced integrated optical setups. Conditional quantum Rabi gates by current hybrid quantum optics technology are directly applicable to extend quantum repeaters for secure quantum optical communication \cite{Repeaters,Repeaters1,Repeaters2,Repeaters3}  
and in future, combine it with error correction strategies \cite{ErrorCorrection,ErrorCorrection1,ErrorCorrection2,ErrorCorrection3,ErrorCorrection4,ErrorCorrection5}. 
It will stimulate further the development of other implementations of the RI and other deterministic nonlinear interactions beyond rotating-wave approximation at the optical frequencies. 

\section*{Acknowledgment}
We thank Adrien Cavailles for a helpful discussion. K.P. acknowledge Project GB19-19722J of the Czech Science Foundation and Danish National Research
Foundation through the Center of Excellence for Macroscopic Quantum States (bigQ, DNRF142). J.L acknowledges Hy-Light project from the French National Research Agency.   R.F. gratefully acknowledges support by the project CZ.02.1.01/0.0/0.0/16 026/0008460 of MEYS CR.

\end{document}